\documentclass[lettersize,journal]{IEEEtran}
\usepackage{amsmath,amsfonts}
\usepackage[ruled,linesnumbered]{algorithm2e}
\usepackage{array}
\usepackage[caption=false,font=normalsize,labelfont=sf,textfont=sf]{subfig}
\usepackage{textcomp}
\usepackage{stfloats}
\usepackage{url}
\usepackage{verbatim}
\usepackage{graphicx}
\usepackage{cite}
\usepackage{multirow, multicol, booktabs, hyperref}
\usepackage{bm,xcolor}
\newcommand{\change}[1]{\textcolor{black}{#1}}

\begin{document}

\title{Curriculum-Based Augmented Fourier Domain Adaptation for Robust Medical Image Segmentation}



\author{An Wang$^\dagger$, Mobarakol Islam$^\dagger$, Mengya Xu$^\dagger$, and Hongliang Ren
\thanks{$\dagger$ equal contribution}
\thanks{This work was supported by the Shun Hing Institute of Advanced Engineering (SHIAE project BME-p1-21) at the Chinese University of Hong Kong (CUHK), Hong Kong Research Grants Council (RGC) Collaborative Research Fund (CRF C4026-21GF and CRF C4063-18G), General Research Fund (GRF 14216022), Shenzhen-Hong Kong-Macau Technology Research Programme (Type C) STIC Grant SGDX20210823103535014 (202108233000303), and (GRS) \#3110167. (Corresponding author: Hongliang Ren.)}
\thanks{A. Wang and H. Ren are with the Department of Electronic Engineering and Shun Hing Institute of Advanced Engineering, The Chinese University of Hong Kong (CUHK), Hong Kong; E-mail: wa09@link.cuhk.edu.hk, hlren@ieee.org; http://labren.org.}
\thanks{M. Xu is with the Department of Biomedical Engineering, National University of Singapore, Singapore, and NUSRI Suzhou, China. E-mail: mengya@u.nus.edu.}
\thanks{M. Islam is with the Wellcome/EPSRC Centre for Interventional and Surgical Sciences (WEISS), Department of Medical Physics and Biomedical Engineering, University College London, London, UK. E-mail: mobarakol.islam@ucl.ac.uk.}}



\maketitle

\begin{abstract}
Accurate and robust medical image segmentation is fundamental and crucial for enhancing the autonomy of computer-aided diagnosis and intervention systems. Medical data collection normally involves different scanners, protocols, and populations, making domain adaptation (DA) a highly demanding research field to alleviate model degradation in the deployment site. To preserve the model performance across multiple \change{testing} domains, this work proposes the Curriculum-based Augmented Fourier Domain Adaptation (Curri-AFDA) for robust medical image segmentation. In particular, our curriculum learning strategy is based on the causal relationship of a model under different levels of data shift in the deployment phase, where the higher the shift is, the harder to recognize the variance. Considering this, we progressively introduce more amplitude information from the target domain to the source domain in the frequency space during the curriculum-style training to smoothly schedule the semantic knowledge transfer in an easier-to-harder manner. Besides, we incorporate the training-time chained augmentation mixing to help expand the data distributions while preserving the domain-invariant semantics, which is beneficial for the acquired model to be more robust and generalize \change{better} to unseen domains. Extensive experiments on two segmentation tasks of Retina and Nuclei collected from multiple sites and scanners suggest that our proposed method yields superior adaptation and generalization performance. Meanwhile, our approach proves to be more robust under various corruption types and increasing severity levels. In addition, we show our method is also beneficial in the domain-adaptive classification task with skin lesion datasets. The code is available at~\url{https://github.com/lofrienger/Curri-AFDA}.
\end{abstract}

\def\abstractname{Note to Practitioners}
\begin{abstract}
Medical image segmentation is key to improving computer-assisted diagnosis and intervention autonomy. However, due to domain gaps between different medical sites, deep learning-based segmentation models frequently encounter performance degradation when deployed in a novel domain. Moreover, model robustness is also highly expected to mitigate the effects of data corruption. Considering all these demanding yet practical needs to automate medical applications and benefit healthcare, we propose the Curriculum-based Fourier Domain Adaptation (Curri-AFDA) for medical image segmentation. Extensive experiments on two segmentation tasks with cross-domain datasets show the consistent superiority of our method regarding adaptation and generalization on multiple \change{testing} domains and robustness against synthetic corrupted data. Besides, our approach is independent of image modalities because its efficacy does not rely on modality-specific characteristics. In addition, we demonstrate the benefit of our method for image classification besides segmentation in the ablation study. Therefore, our method can potentially be applied in many medical applications and yield improved performance. Future works may be extended by exploring the integration of curriculum learning regime with Fourier domain amplitude fusion in the testing time rather than in the training time like this work and most other existing domain adaptation works.

\end{abstract}


\section{Introduction}
\label{sec:introduction}

\begin{figure*}[!hbpt]
\centering
\includegraphics[width=0.88\textwidth]{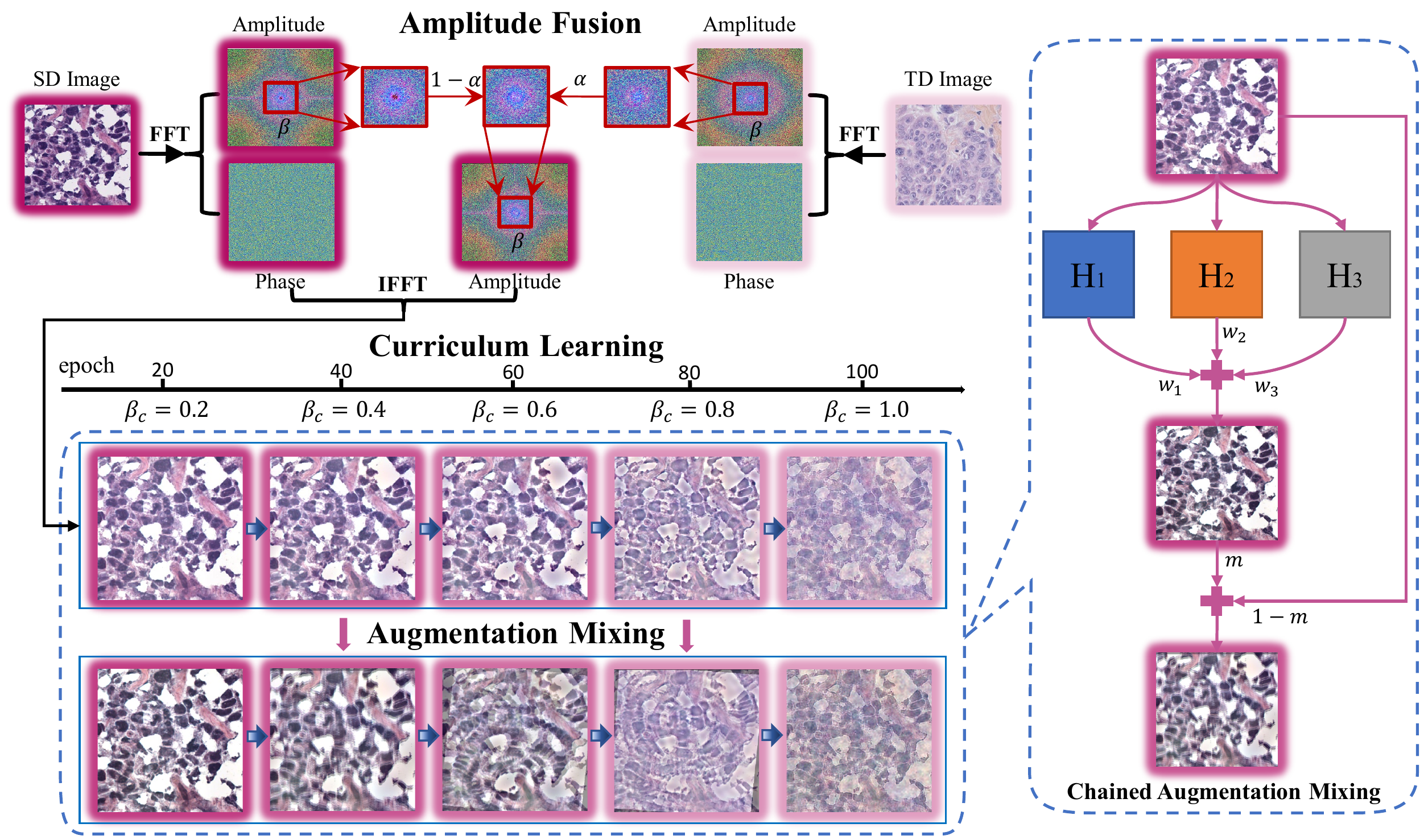}
\caption{\textbf{Overview of the proposed Curriculum-based Augmented Fourier Domain Adaptation (Curri-AFDA).} In the \textbf{Amplitude Fusion} (AF) module, the amplitude scaling coefficient $\beta$ adjusts the central region area of the amplitude spectrum to be mixed between the source domain (SD) and the target domain (TD), and the weighting coefficient $\alpha$ controls the mixing strength. FFT and IFFT stand for the Fast Fourier Transform and the Inverse Fast Fourier Transform. Then the composited images are adopted in the \textbf{Curriculum Learning} (CL) process to train the domain-adaptive model. Amplitude fusion of images gradually gets enhanced when $\beta_c$ linearly grows with epochs, making the source domain data appear more similar to the target domain data. During training, the \textbf{Chained Augmentation Mixing} (CAM) module helps create more variations of the training samples by mixing the outputs of up to three augmentation chains (ACs) and then with the original input image. $H_i$ and $w_i$ represent the sequential augmentations and the mixing weight of the $i^{th}$ chain, respectively. $m$ denotes the mixing weight with the original input.}
\centering
\label{fig:proposal}
\end{figure*}

\IEEEPARstart{A}{lthough} deep learning is showing impressive performance in medical applications to boost the autonomy of computer-aided diagnosis and intervention, recent studies observe significant degradation in the deployed target dataset~\cite{wang2020dofe, liu2021feddg, piva2023empirical}. This is due to  domain shifts such as population shift, covariate shift, and acquisition shift~\cite{rabanser2019failing, castro2020causality, stacke2020measuring} in the deployment domain. In particular, the problem is usually unavoidable in the medical imaging field because medical data and annotations are usually limited and derived from multiple working sites with different scanners, protocols, and populations. This problem also leads to overfitting, underspecification, poor generalization and weak robustness of the model.

Many works have focused on domain adaptation (DA) and domain generalization (DG) to tackle data shifts in the target domain. Most of these works utilize supervised, semi-supervised, and unsupervised techniques with the strategies of transfer learning~\cite{kocielnik2023can, cao2019learning}, fine-tuning~\cite{valverde2019one, chambon2023improved}, adversarial training~\cite{haq2020adversarial, xu2021learning, vu2019advent}, and data augmentation~\cite{wang2023domain, brion2021domain}. Depending on the availability of the target domain data during training, there are typically two types of domain adaptation. Testing-time DA, like domain generalization, tries to handle unseen domain shifts from training. Training-time DA, where target domain data is available with limited annotations (weakly-supervised DA) or no annotations (unsupervised DA), mainly emphasizes transferring target domain information to the source domain during training. The transferring methods can also be categorized as feature-level transferring~\cite{zhang2017curriculum, luo2019significance}, image-level transferring~\cite{jiang2020psigan, chen2019synergistic}, and label-level transferring~\cite{xia2020uncertainty}. Recently, Fourier Transform has been used to transfer domain-specific information from target images to source images by performing amplitude fusion in the frequency domain~\cite{yang2020fda, xu2021fourier, liu2021feddg, yang2021hcdg, yang2022source}. These studies show the effectiveness of the Fourier technique with the advantage of simplicity and model-agnostic characteristics. However, besides adaptation and generalization, the above works seldom explore robustness under naturally-induced data alterations and corruption, which is also crucial in the model deployment phase.

\change{Recently, curriculum learning~\cite{bengio2009curriculum}, a training scheme that aims to let the model learn from easier to more complex samples or tasks, has been captivating increasing attention in the field of computer vision. One of the key benefits of curriculum learning is that it can improve a model's generalization performance.}
The \change{efficacy} of replacing conventional training with curriculum learning has been demonstrated in many application fields, such as semantic segmentation~\cite{islam2023paced, zhang2017curriculum}, object detection~\cite{sangineto2018self, yang2023dynamic}, neural machine translation~\cite{kocmi2017curriculum}, image captioning~\cite{xu2021class}, and robotic learning~\cite{portelas2020teacher}. The efficacy of a curriculum-based model mainly depends on the proper design of the difficulty measurement process for training samples or tasks. Specifically for curriculum-based Domain Adaptation, different approaches such as domain discriminator~\cite{yang2020curriculum}, density-based clustering~\cite{choi2019pseudo}, superpixel label transfer~\cite{zhang2017curriculum}, and domain similarity grouping~\cite{zhang2019curriculum} have been proposed to quantify difficulty in a weakly supervised or unsupervised manner.

The generalization and robustness abilities of the deep learning model are frequently observed to be improved by augmenting training data. Multiple augmentation techniques have been developed to boost model performance in a cutting-based or mixing-based manner, e.g., CutOut~\cite{devries2017improved}, MixUp~\cite{zhang2018mixup}, CutMix~\cite{yun2019cutmix}, and AugMix~\cite{hendrycks2020augmix}. To assess the robustness of the deep learning model, benchmark datasets for two types of robustness (corruption and perturbation) are created~\cite{hendrycks2019benchmarking}. The enhanced robustness is demonstrated and proven with an altered test dataset that includes corrupted and perturbed images~\cite{geirhos2019imagenet, zhang2019making}.

In this work, we design a Curriculum-based Augmented Fourier Domain Adaptation (Curri-AFDA) method to tackle domain shift by transferring target domain information to the source domain in a curriculum manner and extensively augmenting the data by training-time chained augmentation mixing, as shown in Fig.~\ref{fig:proposal}. To build the curriculum strategy, we consider modeling the difficulty of domain adaptation as recognizing target domains with different levels of distribution shift. Specifically, we utilize Fourier Transform to extract and fuse the source and target domain information over the training period in an easier-to-harder curriculum order by progressively increasing the amplitude transferred from the target to the source domain in the frequency space. Then the reconstituted training samples are passed through chains of various augmentation operations in random orders to further improve data diversity. We validate the proposed approach on two medical image segmentation datasets of Retina and Nuclei segmentation collected from multiple domains with obvious domain shifts. We also evaluate the robustness of our method by applying 15 different corruption and perturbation techniques with five increasing severity levels on the test dataset. Extensive cross-domain validation and robustness results suggest that our approach not only improves the performance of mitigating domain variance but is also highly robust against heavy data corruptions.

Our main contributions and findings can be summarized as follows:
\begin{itemize}
\item Demonstrate the progressively incremental amplitude fusion in the Fourier space as an effective curriculum-based approach to alleviate domain discrepancy.
\item Incorporate the training-time chained augmentation mixing to further boost the training data diversity and establish the Curriculum-based Augmented Fourier Domain Adaptation (Curri-AFDA).
\item Conduct extensive experiments on multiple Retina segmentation and Nuclei segmentation datasets and various types and levels of corrupted datasets to show the superiority of our method with respect to adaptation, generalization, and robustness. 
\item Explore the efficacy of Curri-AFDA for the image classification task besides segmentation with the skin lesion datasets and show the potential of our method for broader medical applications.
\end{itemize}

\section{Related works}

\subsection{Fourier Transform for Domain Adaptation} 
Due to its simplicity, effectiveness, and model-agnostic characteristic, Fourier Transform is one of the recent tools in Domain Adaptation. In the Fourier-based frequency space, the low-frequency amplitude components, i.e., the central region of the amplitude spectrum, carry more domain-specific information. Fourier Domain Adaptation (FDA)~\cite{yang2020fda} applies Fourier Transform and its inverse to spatial images and fuses the amplitude spectrum in the low-frequency region of the source domain and target domain samples to tackle domain shift. Similarly, amplitude fusion is performed by preserving the phase information for unsupervised domain adaptation~\cite{yang2020phase}. Basically, the Fourier-based domain adaptation method tries to mitigate the domain gap by image-to-image translation (I2I) or style transfer - one of the major strategies for domain adaptation. After style transfer, the source domain data is expected to share a similar style as the target domain. By amplitude mixing in the frequency space, the training data appears to be in an intermediate style between the SD and TD, depending on the fusion strength. At the same time, the core domain-invariant semantics information remains unchanged in the generated image. Instead of swapping low-frequency amplitude components, Fourier augmented co-teacher (FACT)~\cite{xu2021fourier} and AmpMix~\cite{xu2023fourier} proposes to mix the whole amplitude spectrum with the MixUp~\cite{zhang2018mixup} technique and achieves better generalization ability. By assigning pixel-wise significance with Gaussian distribution and introducing pixel-wise disturbance in the amplitude spectrum, HCDG~\cite{yang2021hcdg} proposes to highlight the core information in the center area of the image than the marginal area. Moreover, in the federated learning scenario, Federated Domain Generalization (FedDG)~\cite{liu2021feddg} constructs a continuous frequency space, where low-frequency amplitude components from multiple remote domains/sites are extracted, stored, and used for training. 

Compared with GAN-based domain adaptation methods, Fourier-based methods avoid additional efforts of complicated adversarial training to accomplish the domain alignment. Besides, in the case of limited training data, GAN-based methods may fail to work since they are known to be heavily data-hungry. Whereas, Fourier-based approaches are less affected and thus have significant advantages in resolving domain shift problems with insufficient data, which is meaningful for practical medical applications.

The methods mentioned above all have a fixed amplitude fusion process. For example, the portion of amplitude components to be transferred from the target domain image to the source domain image remains the same throughout the entire training. On the contrary, our proposed method introduces more target domain information to the source domain by progressively increasing the number of mixed amplitude components following the training scheduling functions. In this way, we implement a curriculum-style dynamic training scheme for the frequency domain adaptation and generate more variations of the training data in a ``easier to harder" order.

\subsection{Curriculum-based Domain Adaptation}

The curriculum-style domain adaptation approach has attracted the interest of the research community due to its excellent generalization ability. The core idea of this strategy is to learn ``from easier to harder" either from the perspective of tasks or samples. At the task level, the work~\cite{stretcu2021coarsetofine} designs the curriculum in a coarse-to-fine manner by decomposing challenging tasks into sequences of easier intermediate goals that are used to pre-train a model before tackling the target task. The efficacy of a curriculum scheme mostly depends on the appropriate difficulty measurement process. Various techniques are utilized for this purpose in domain adaptation tasks. For example, a domain discriminator to measure easier domain for multi-source domain adaptation~\cite{yang2020curriculum}, a density-based clustering algorithm to sort the samples from the target domain based on distance~\cite{choi2019pseudo}, and semantically easier class region can be considered as the easier label to train in curriculum strategy~\cite{zhang2017curriculum}. 

Unlike previous works, we apply curriculum-based domain adaptation by gradually introducing domain-variant information from the target domain (TD) to the source domain (SD) to mitigate the domain shift. Specifically, we take advantage of the property that the amplitude components of the frequency-domain image contain essential and specific low-level statistics. Then we design a progressive style alignment method between the source domain and the target domain by amplitude fusion of images. In such a manner, the training samples will carry more target domain information and appear more 
similar to the target domain images in the later training phase. \change{By building up understanding slowly and systematically through our carefully designed curriculum, the model is able to learn more robust, generalized representations of the data. This can lead to improved performance on new, unseen data, as the model has learned to recognize more complex patterns and generalize them to new situations.}

\subsection{Data Augmentation by Mixing Images}
To overcome the problem of overfitting, poor robustness, and weak generalization of deep learning models, various approaches have been proposed. Among them, data augmentation techniques, which create novel variations of the existing training images, have gained continuous attention over the years. Apart from traditional techniques like color mutation and geometric transformation, data augmentation can also be done by simply removing part of the original image~\cite{devries2017improved} or further replacing it with a certain noise~\cite{yun2019cutmix}. Except for cutting, another line of research also apply image mixing to generate new images. A pioneer mixing method is MixUp~\cite{zhang2018mixup}, followed by many other works in this area~\cite{huang2021snapmix, hendrycks2020augmix, liu2023decoupled}. Among them, AugMix~\cite{hendrycks2020augmix} is different in that it mixes more than two images from up to three augmentation chains. In each augmentation chain, several base augmentation operations (e.g., translation, rotation, auto-contrast) are arbitrarily applied to the original image. Then the augmented images from all chains are linearly mixed with the original image to form an overall training sample. \change{The use of mixing augmentation can enhance the diversity of training data, which is crucial for improving robustness against unexpected shifts and corruptions in data.}

Considering this, we also design the chained augmentation mixing strategy in our curriculum-based training process to enhance the generalization and robustness performance. Compared with sequentially conducting separate augmentations in a normal training scheme, our one-step chained augmentation mixing is more efficient in improving the training data diversity, only with minimal cost of matrix-weighted addition and no other computational complexity in neither the training nor test stage.

\section{Methodology}
\label{sec:method}
In this work, we design a curriculum-based cross-domain information fusion strategy in the Fourier space and incorporate the training-time chained augmentation mixing module to improve the model performance concerning adaptation, generalization, and robustness against natural and synthetic data shifts. As shown in Fig.~\ref{fig:proposal}, our method mainly consists of three components: Amplitude Fusion, Curriculum Learning, and Chained Augmentation Mixing.

\subsection{Amplitude Fusion in the Fourier Space}
For a spatial-domain digital image $x$, we can extract the amplitude components $A(x)$ and the phase components $P(x)$ in the frequency domain with the Fourier Transform of $x$, i.e., $F(x)$. As the amplitude components of the Fourier Transform carry the most domain-specific information~\cite{oppenheim1981importance, yang2020phase, piotrowski1982demonstration, guyader2004image}, for Domain Adaptation (DA), most frequency-domain image processing techniques manipulate only the amplitude spectrum while preserving the phase spectrum as it is critical for maintaining the overall visual look of an image~\cite{oppenheim1981importance}. The pioneering work FDA~\cite{yang2020fda} attempts to tackle the domain shift problem by mutating the center region of $A(x)$ from the source domain (SD) with that from the target domain (TD) in the frequency space. If $A_S$ and $A_T$ are denoted as the amplitude components of two random images from SD and TD, the reconstituted amplitude components in the frequency space $A_S^F$ at the point $(u,v)$ can be formulated as-
\begin{equation}
\resizebox{0.91\hsize}{!}{$
    A_S^F(u,v) =
    \begin{cases} 
    (1-\alpha)A_S(u,v)+\alpha A_T(u,v), &\text{if }u, v \in [-\hat \beta, \hat \beta]\\
    A_S(u,v), &\text{otherwise}
    \end{cases}
    $}
    \label{eq:fda}
\end{equation}
where $\hat \beta = ||\beta H||$ or $\hat \beta = ||\beta W||$ for $u$ and $v$ respectively and $||$ is the floor rounding operation. $H$ and $W$ are the height and width of the image. The weighting coefficient $\alpha \in[0,1]$ controls the mixing ratio of amplitude components from $A_S$ and $A_T$. The amplitude scaling coefficient $\beta \in[0,1]$ adjusts the area of the mutated center region and a larger $\beta$ means a larger center region of $A_S$ and $A_T$ will be used for amplitude fusion. Eventually, with the inverse Fourier Transform $F^{-1}$, the reconstituted spatial-domain SD image $x_S^F$ can be expressed as $x_S \rightarrow x_S^F=F^{-1}(A_S^F, P_S)$. Both Fourier Transform and its inverse can be efficiently implemented by the FFT~\cite{nussbaumer1981fast} algorithm.

\subsection{Chained Augmentation Mixing}
Data augmentation can significantly increase generalization and robustness performance by introducing a higher diversity in training data. Furthermore, by stochastically sampling and mixing various augmentation methods with the original image, we can generate more novel augmented images without deviating too far from the original. Varieties of augmentation operations are covered in the augmentation chains (ACs), such as auto-contrast, equalization, posterization, rotation, solarization, shear, and translation in serial and parallel orientations. For a spatial-domain image $x$, after the chained augmentation mixing, the augmented image $x^{Aug}$ can be expressed as below,
\begin{equation}
    x^{Aug} = m \cdot x + (1-m) \cdot \sum_{i=1}^{AC}{(w_i\cdot H_i(x))}
    \label{eq:augmix}
\end{equation}
where $m$ is a random convex coefficient sampled from a Beta distribution \change{$B(\cdot)$}, $w_i$ is another random convex coefficient sampled from a Dirichlet distribution \change{$D(\cdot)$} controlling the mixing weights of the augmentation chains, and $H_i$ denotes the sequential augmentation operations applied to the $i^{th}$ augmentation chain. Each augmentation chain consists of up to three base augmentation operations that are chosen at random. Details are illustrated in the right part of Fig.~\ref{fig:proposal}.

\subsection{Curri-AFDA: Curriculum-based Augmented Fourier Domain Adaptation}
\begin{figure}[!hbpt]
\includegraphics[width=0.95\columnwidth]{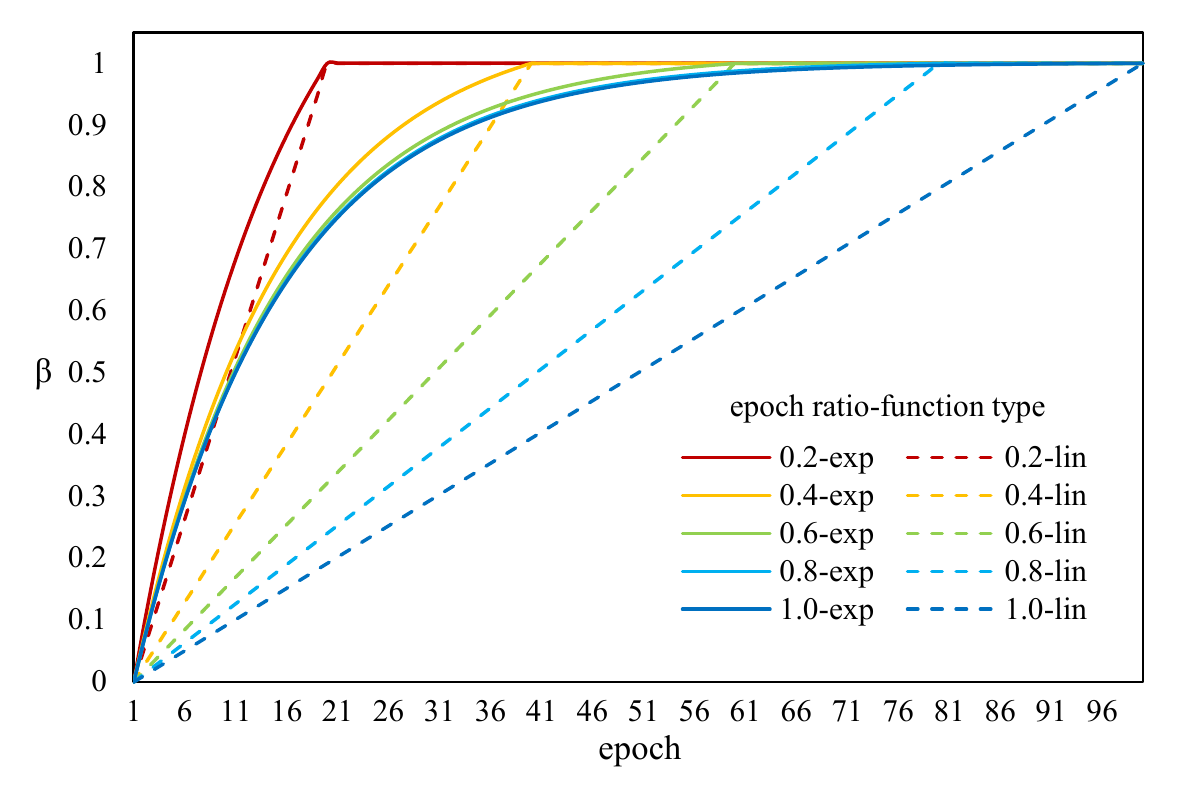}
\centering
\caption{\textbf{Visualization of the linear and exponential increment of amplitude scaling coefficient ($\beta$) with different epoch ratios.} Different line colors indicate different epoch ratios, and two line types differentiate two scheduling functions.}
\label{fig:lin_exp}
\end{figure}

For a model trained on a single domain data, it is easier to recognize images from the same domain and harder from another domain with data shift. In our curriculum strategy, the amplitude components from TD are progressively transferred to SD in the frequency space over the training period. In this way, the model learns comparatively easier information first from a single domain and successively adopts harder features like changes in the distribution of the input data from other domains. More specifically, we control the effect of the amplitude fusion by gradually increasing the amplitude scaling coefficient $\beta$ from 0 to the optimal value $\beta_{opt}$. As $\beta$ grows, the reconstituted training samples will gradually carry more target domain information, letting the model learn the distribution changes for the target domain. Besides, because this fusion process is slight in the early training phase, the model could firstly focus on source domain samples to recognize domain-invariant basic features without being affected by aggressive target domain information.

\change{To facilitate our strategy that transforming the training data in each epoch following the curriculum order, i.e., ``easier to harder'' or ``cleaner to noisier'', we first employ a linearly increasing scheduler function.} Specifically, if $\beta_c$ is the scaling coefficient in the curriculum stage, then the linear scheduling function can be formulated as-
\begin{equation}
    \beta_c =
    \begin{cases} 
    \frac{e}{E\cdot r_e}\cdot \beta_{opt}, &\text{if }e \leq E\cdot r_e\\
    \beta_{opt}, &\text{otherwise}
    \end{cases}
    \label{eq:lin_sche}
\end{equation}
where $E$ is the total number of training epochs, $e$ is the current epoch, $r_e$ stands for epoch ratio which controls the length of the curriculum stage in the complete training stage and further controls the changing rate of $\beta_c$ with a fixed optimal scaling coefficient $\beta_{opt}$. As the epoch $e$ grows, $\beta_c$ also increases, resulting in incremental amplitude mixing as indicated in~\eqref{eq:fda}. In this progress, the model is gradually exposed to more target domain-specific information to improve adaptive ability continuously. 
\change{Besides the linear scheduler function, there are several other candidates used in Curriculum Learning to provide distinctive learning paths. We also try the exponential scheduler function, as depicted in Fig.~\ref{fig:lin_exp}.}

As a result, instead of using constant or random $\beta$ in other Fourier-based adaptation methods, we adopt the incremental $\beta_c$ and reconstitute the new training sample $x_S^{CF}$ with the inverse Fourier Transform which can be represented as-
\begin{equation}
    x_S^{CF}=F^{-1}(A_{S(\beta_c)}^F,P_S).
    \label{eq:cfda}
\end{equation}
These generated images are then fed into the chained augmentation mixing module to produce more variations of the training data. Through this, we can improve the training data diversity further and thus boost the generalization and robustness performance of the model. The final reconstituted training image, $x_S^{CAF}$, can thus be expressed according to~\eqref{eq:augmix} as-
\begin{equation}
    x_S^{CAF} = m \cdot x_S^{CF} + (1-m) \cdot \sum_{i=1}^{AC}{(w_i\cdot H_i(x_S^{CF}))}.
\label{eq:caf}
\end{equation}

The perturbation in the low-frequency amplitude components of an image in the Fourier space will not alter the core semantics of the original image, such as the Nuclei shapes. Therefore, the masks remain unchanged in the cross-domain amplitude fusion process. Whereas in case of geometric changes during the chained augmentation mixing, the same transformations are applied to both the images and masks to adjust with the shape deviation.

\SetKwComment{Comment}{/* }{ */}
\begin{algorithm}[!t]
    \caption{\change{Pseudo code of Curri-AFDA.}}
    \label{alg:proposed_algo}    
    \change{\KwIn{Source/target domain image $x_S$/$x_T$, current/total epoch $e$/$E$, epoch ratio $r_e$, current/optimal amplitude scaling coefficient $\beta_c$/$\beta_{opt}$, amplitude mixing coefficient $\alpha$.}}
    \change{\KwOut{Final reconstituted image $x_S^{CAF}$.}}
    \change{Initialize $\alpha, \beta_{opt}, r_e, E$\;}
    \change{
    \While{$e < E$}{
        \tcp{Get Amplitude ($A$) and Phase ($P$) by Fourier Transform ($F$)}
        $A_S, P_S \leftarrow F(x_S); A_T, P_T \leftarrow F(x_T)$\;
        \tcp{Scheduling scaling coefficient}
        \eIf{$e \leq E \cdot r_e$}{  \label{alg:sche}
        $\beta_c \leftarrow scheduler(\beta_{opt})$\;
        }{
        $\beta_c \leftarrow \beta_{opt}$\;
        }
        \tcp{Amplitude fusion}
        $A_S^F \leftarrow AF(A_S, A_T) \text{ s.t. } \alpha, \beta_c$\; \label{alg:AF}
        \tcp{Inverse Fourier Transform}
        $x_S^{CF} \leftarrow F^{-1}(A_S^F, P_S)$\;
        \tcp{Chained augmentation mixing}
        $x_S^{CAF} \leftarrow CAM(x_S^{CF}) \text{ s.t. } m \sim B(\cdot), w_i \sim D(\cdot)$\; \label{alg:CAM}
        Take $x_S^{CAF}$ as input for training.
    }}
\end{algorithm}

Until this, we have elaborated our \textbf{Curri}culum-based \textbf{A}ugmented \textbf{F}ourier \textbf{D}omain \textbf{A}daptation (Curri-AFDA), a novel approach to resolve model degradation in case of domain shifts and data corruptions.
\change{Algorithm~\ref{alg:proposed_algo} outlines the pseudo code to implement our proposed method efficiently. In the training process, we employ Fourier Transform to acquire the amplitude components of both SD and TD images. The resultant scaling coefficient, generated by the scheduler function, regulates the amplitude fusion process. Next, the Inverse Fourier Transform is applied to produce a new image. Subsequently, the application of chained augmentation mixing facilitates the generation of additional image variants.} 
It is worth mentioning that only one image is generated from each source-target image pair in every epoch, so there is no additional training memory consumption. However, owing to the curriculum-based cross-domain information fusion in the Fourier space, the training data distribution gradually becomes closer to the target domain. The diversity of the training data also gets boosted by the chained augmentation mixing module, which is beneficial to make the model more generalizable and robust. 

\section{Experiments}
\label{sec:exp}

\subsection{Datasets} 
\begin{figure}[!t]
\centering
\includegraphics[width=\columnwidth]{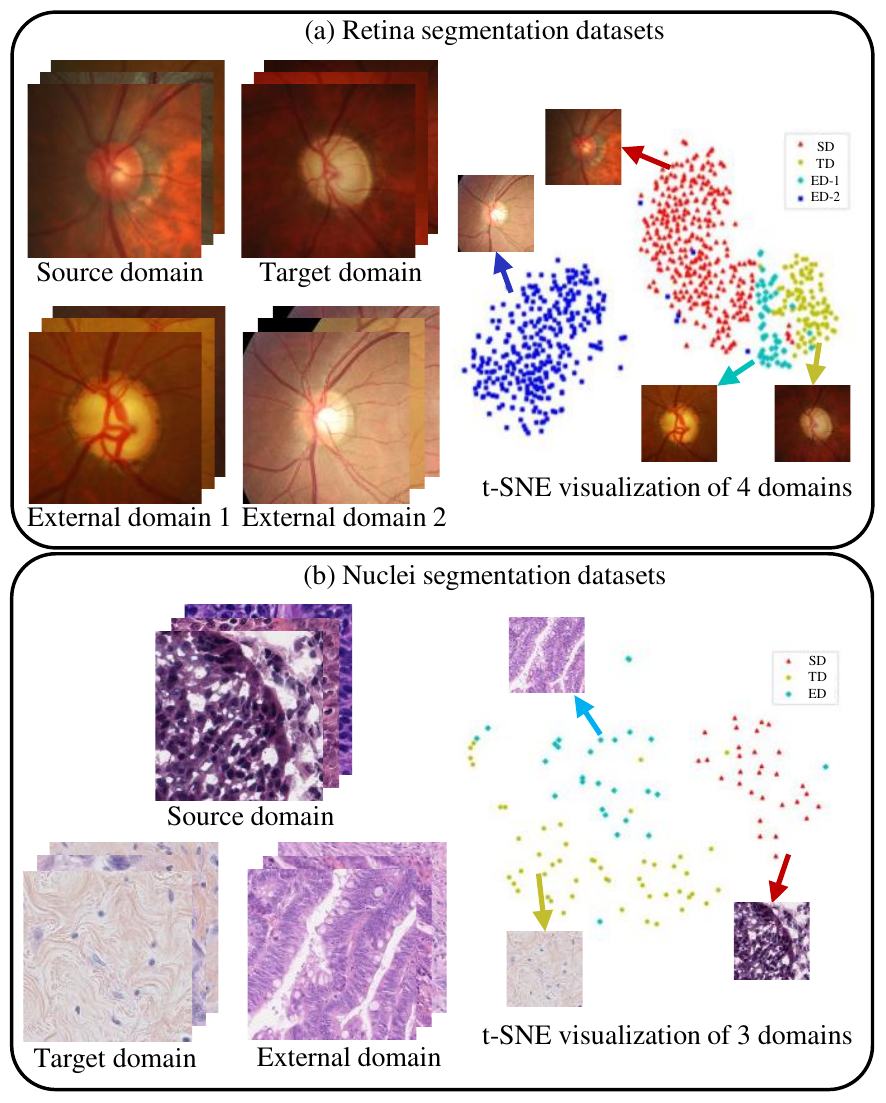}
\caption{\textbf{Example images of source and \change{testing} domains in (a) Retina segmentation and (b) Nuclei segmentation tasks.} The t-SNE visualization of image features (extracted by a ResNet-101 network pre-trained on ImageNet) indicates a significant domain shift.}
\label{fig:dataset}
\end{figure}

\change{We perform extensive validation of our method on two widely-used and well-established medical image segmentation benchmark tasks, i.e., Retina optic cup and disc segmentation on fundus images~\cite{wang2020dofe,liu2021feddg,zhou2022generalizable,lei2021unsupervised} and Nuclei segmentation~\cite{haq2020adversarial,sharma2022mani,li2022domain}. The Retina and Nuclei databases are collected from different imaging modalities, where Retina and Nuclei images are collected from color fundus photography (CFP) and pathological scanning, respectively. Besides, they comprise four and three data sources, featuring typical domain shifts such as imaging resolution, data quality, and patient populations. Therefore, they can facilitate the model assessment regarding adaptation, generalization, and robustness.}
\change{For every segmentation task, we assign one domain as the source domain (SD) and select another domain as the target domain (TD) for amplitude fusion during training. The remaining domains are treated as external, unseen domains that are only used to test generalization and robustness.}

\textbf{Retina segmentation datasets} are collected from  four different scanners and sources, i.e., Drishti-GS~\cite{sivaswamy2015comprehensive}, RIM-ONE-r3~\cite{fumero2011rim}, REFUGE-train~\cite{orlando2020refuge}, and REFUGE-valid~\cite{orlando2020refuge}. There are two annotation labels of the optic disc and optic cup for all the datasets. 
\change{These datasets are collected and pre-processed by DoFE~\cite{wang2020dofe} in their domain generalization task. Here we employ the database in a single-source setup containing one fixed source domain (SD), one fixed target domain (TD), and two external domains (EDs).}
\change{REFUGE-train~\cite{orlando2020refuge}, RIM-ONE-r3~\cite{fumero2011rim}, Drishti-GS~\cite{sivaswamy2015comprehensive}, and REFUGE-valid~\cite{orlando2020refuge} have 400, 159, 101, and 400 samples, respectively. To learn more general features and mitigate potential model bias resulting from a small training data size, we designate the REFUGE-train dataset as the source domain (SD) and the remaining datasets as the target domain (TD) and two external domains (ED-1 and ED-2). Additionally, while the REFUGE-train and REFUGE-valid datasets have an identical size of 400, we maintain their original train-valid split~\cite{orlando2020refuge} for the purposes of training and testing without any modifications.}
Fig.~\ref{fig:dataset}(a) presents some random samples from each domain and the corresponding embedded feature representation. A clear domain shift can be observed from both the appearance and the embedding space.

\textbf{Nuclei segmentation datasets} are collected from three sources where CryoNuSeg~\cite{mahbod2021cryonuseg} is treated as the source domain while TNBC~\cite{naylor2018segmentation} and CoNSeP~\cite{graham2019hover} are \change{the target domain (TD) and external domain (ED)}. CryoNuSeg~\cite{mahbod2021cryonuseg} dataset is extracted from the Cancer Genome Atlas (TCGA). It contains 30 images collected from 10 different human organs (three images per organ), namely the adrenal gland, larynx, lymph node, mediastinum, pancreas, pleura, skin, testis, thymus, and thyroid gland. TNBC (Triple Negative Breast Cancer)~\cite{naylor2018segmentation} dataset is acquired at Curie Institute, containing 50 images from 11 patients. The CoNSeP~\cite{graham2019hover} dataset consists of 41 images, including stroma, glandular, muscular, collagen, fat and tumour regions. The data from TNBC~\cite{naylor2018segmentation} is used for cross-domain information fusion. It is randomly split for training and testing with a ratio of 8:2, similar to the dataset split strategy introduced in~\cite{weiss2020processing, sharghi2020automatic}. The testing split of \change{TNBC~\cite{naylor2018segmentation} and the entire CoNSeP~\cite{graham2019hover} dataset} are not accessed during training. The domain shift between these datasets arises from organ differences, institutional differences, and different imaging tools and protocols. Obvious domain gaps are visualized in Fig.~\ref{fig:dataset}(b) with the t-SNE embedding feature representations.

\subsection{Implementation Details}
We implement our method on top of a state-of-the-art segmentation backbone, UNet~\cite{ronneberger2015unet} and a recent Swin-Transformer-based model Swin-UNet~\cite{cao2023swin}. A vanilla UNet architecture~\footnote{https://github.com/ternaus/robot-surgery-segmentation} and the official Swin-UNet implementation~\footnote{https://github.com/HuCaoFighting/Swin-Unet} with the pretrained Swin-Transformer~\cite{liu2021swin} weight~\footnote{https://github.com/microsoft/Swin-Transformer} are adopted. The images are resized to 384$\times$384 and 224$\times$224 for UNet~\cite{ronneberger2015unet} and Swin-UNet\cite{cao2023swin} models, respectively. In addition to the aforementioned Fourier-based FDA~\cite{yang2020fda} and FACT~\cite{xu2021fourier}, we also take the adversarial-based segmentation method ADVENT~\cite{vu2019advent} as another baseline. We refer to the official repository~\footnote{https://github.com/valeoai/ADVENT} for implementation, such as the discriminator model and its hyper-parameters. The Fourier parts in our proposal are realized with the official implementation of FFT (Fast Fourier Transform) and IFFT (Inverse Fast Fourier Transform) from the Python Numpy library. 

In the curriculum stage, the optimal amplitude scaling coefficient $\beta_{opt}$ is firstly determined from the vanilla FDA~\cite{yang2020fda} by empirically tuning and deriving the best constant scaling coefficient. Eventually, the value of $\beta_{opt}$ is 0.006 and 1.0 in the Retina and Nuclei segmentation. Then according to~\eqref{eq:lin_sche}, we schedule $\beta_c$ by various epoch ratios $r_e$ and the fixed $\beta_{opt}$. For a fair comparison, we keep the weighting coefficient $\alpha$ constant for all experiments. Specifically, for the UNet~\cite{ronneberger2015unet} backbone, $\alpha$ is set as 1.0 and 0.7 in the Retina and Nuclei Segmentation, while for the Swin-UNet~\cite{cao2023swin} backbone, $\alpha$ is 0.5 and 0.7 in the two tasks. Further details of parameter tuning are presented and discussed in the ablation study section~\ref{sec:ablation}. 

For the augmentation mixing, we modify the official implementation~\footnote{https://github.com/google-research/augmix} of AugMix~\cite{hendrycks2020augmix} to adapt it to our curriculum-based amplitude fusion training process. The augmentation level, which controls the transformation strength globally, is set as 3 and 2 in the UNet-based and Swin-UNet-based backbones. The number of augmentation chains (ACs) is 3 and each chain includes up to 3 stochastically sampled transformations. The hyperparameters in the Beta and Dirichlet distribution are all set as 1. In addition, we use a learning rate of 0.001 and the Adam optimizer for training.

To evaluate the robustness of other methods and our Curri-AFDA, we adopt various corruption techniques to construct a series of synthetic Retina datasets. Specifically, four groups of corruptions, i.e., noise, blur, weather, and digital, including 15 corruption operations, i.e., ``Gaussian, Shot, Impulse", ``Defocus, Glass, Motion, Zoom", ``Snow, Frost, Fog, Bright", and ``Contrast, Elastic, Pixel, JPEG Compression" are utilized to generate the test datasets. Furthermore, each type of corruption has five levels of severity. In this way, we can thoroughly assess the robustness under various corruption types and levels.

\begin{table*}[!htbp]
\centering
\caption{\textbf{Quantitative results on Retina segmentation and Nuclei segmentation.} \change{For both tasks, only the target domain (TD) is adopted for the amplitude fusion with the source domain (SD) during training. The \change{external domains} are used during testing to evaluate the generalization robustness.} DSC (\%) is adopted as the performance metric. Average results across all \change{testing} domains and the corresponding standard deviations (STD) are presented for both tasks. The vanilla method means no domain adaptation approach is applied. The best results are shown in bold and the runner-up results are underlined.}
\resizebox{0.88\textwidth}{!}{%
\begin{tabular}{c|ccccc|cccc} 
\toprule
\multirow{2}{*}{\textbf{Methods}}   & \multicolumn{5}{c|}{\textbf{Retina Segmentation}}                                                               & \multicolumn{4}{c}{\textbf{Nuclei Segmentation}}                                                                                                 \\ 
\cline{2-10}
                                     & $\beta$                                                                        & TD           & \change{ED-1}           & \change{ED-2}           & Mean$\pm$STD                    & $\beta$                                                            & TD           & \change{ED}           & Mean$\pm$STD                     \\ 
\hline
Vanilla UNet~\cite{ronneberger2015unet}                                                                     & N.A.                                                                        & 70.33          & 75.52          & 61.65          & 69.17$\pm$7.01                  & N.A.           & 13.17          & 9.71           & 11.44$\pm$2.45                   \\
                                     + ADVENT~\cite{vu2019advent}                                                                       & N.A.                                                                        & 77.10          & 71.54          & 66.10          & 71.58$\pm$5.50                  & N.A.         & 27.42          & 28.09          & 27.76$\pm$0.34           \\
                                     + FDA~\cite{yang2020fda}                                                                          & 0.006                                                                      & \underline{79.48}  & 76.55          & \underline{85.74}  & 80.59$\pm$4.70                  & 1                                      & 44.19          & 41.70          & 42.95$\pm$1.76                   \\
                                     + FACT~\cite{xu2021fourier}                                                                         & 1                                                              & 78.18          & 76.29          & 83.00          & 79.16$\pm$3.46                  & 1                                                                & 33.33          & 32.61          & 32.97$\pm$0.36                   \\ 
\hline
                                     \multirow{3}{*}{+ Curri-AFDA (Ours)} & \begin{tabular}[c]{@{}c@{}}linear\\ (0 to 0.006)\end{tabular}       & 78.85          & \textbf{83.15} & 84.59          & \underline{82.20}$\pm$2.99 & \begin{tabular}[c]{@{}c@{}}linear\\ (0 to 1)\end{tabular}          & \underline{46.02}  & \underline{46.29}  & \underline{46.16}$\pm$0.14  \\
                                     & \begin{tabular}[c]{@{}c@{}}exponential\\ (0 to 0.006)\end{tabular}          & \textbf{80.44} & \underline{80.79}  & \textbf{85.89} & \textbf{82.37}$\pm$3.05 & \begin{tabular}[c]{@{}c@{}}exponential\\ (0 to 1)\end{tabular}        & \textbf{50.61} & \textbf{52.29} & \textbf{51.45}$\pm$0.84          \\ 
\midrule
Vanilla Swin-UNet~\cite{cao2023swin}                                                                     & N.A.             & 76.89          & 72.57          & 84.33          & 77.93$\pm$5.95                  & N.A.                                   & 23.95          & 20.73          & 22.34$\pm$1.61          \\
                                     + ADVENT~\cite{vu2019advent}                                                                       & N.A.                        & 76.35          & 74.88          & \underline{86.97}  & 79.40$\pm$6.60                  & N.A.                                                      & 40.09          & \textbf{48.57} & 44.33$\pm$4.24                   \\
                                     + FDA~\cite{yang2020fda}                                                                          & 0.006                                           & 77.89          & \textbf{77.83} & 70.98          & 75.57$\pm$3.97                  & 1                                             & 55.09          & 40.40           & 47.75$\pm$7.35                   \\
                                     + FACT~\cite{xu2021fourier}                                                                         & 1          & 76.74          & 75.36          & 81.55          & 77.88$\pm$3.25         & 1                                                                 & 52.54          & \underline{45.68}  & 49.11$\pm$3.43           \\ 
\hline
\multirow{3}{*}{+ Curri-AFDA (Ours)} & \begin{tabular}[c]{@{}c@{}}linear\\ (0 to 0.006)\end{tabular}   & \textbf{83.19} & \underline{77.81}  & 85.24          & \textbf{82.08}$\pm$3.84 & \begin{tabular}[c]{@{}c@{}}linear\\ (0 to 1)\end{tabular}            & \underline{60.62}  & 43.03          & \underline{51.83}$\pm$8.80           \\
                                     & \begin{tabular}[c]{@{}c@{}}exponential\\ (0 to 0.006)\end{tabular}   & \underline{78.64}  & 73.31          & \textbf{88.42} & \underline{80.12}$\pm$7.66          & \begin{tabular}[c]{@{}c@{}}exponential\\ (0 to 1)\end{tabular}         & \textbf{61.75} & 45.04          & \textbf{53.40}$\pm$8.36          \\
\bottomrule
\end{tabular}
\label{tab:overall_result}
}
\end{table*}

\section{Evaluation and Results}
\label{sec:eva_res}
\subsection{Evaluation Description}
To evaluate the segmentation performance, we use a commonly-used metric, Dice Similarity Coefficient (DSC). \change{We also compute the mean and standard deviations of the results for all testing datasets to present the overall model performance}. The performance of our method is compared with two closely related works, i.e., FDA~\cite{yang2020fda} and FACT~\cite{xu2021fourier}, on top of the vanilla CNN-based UNet~\cite{ronneberger2015unet} and Transformer-based Swin-UNet~\cite{cao2023swin}. Besides, the GAN-based method ADVENT~\cite{vu2019advent} is also adopted as another reference baseline. 
\change{We have conducted extensive assessments of our method across 1) domain-adaptive performance on the target domain, 2) generalization ability on previously unseen domains, and 3) robustness to both natural and synthetic data shifts and corruptions. Our evaluation settings follow the standard unsupervised domain adaptation (UDA)~\cite{yang2020fda}, the generic specification of model generalization~\cite{zhou2022domain} and external validity~\cite{eertink2022external}, and the benchmark assessment of robustness~\cite{hendrycks2019benchmarking}.}

In real medical applications, due to data scarcity, deep learning models are often required to handle different unseen data shifts to achieve testing-time adaptation. Considering this, we not only evaluate the training-time DA performance with TD, which is available for amplitude fusion during training, but also perform the testing-time evaluation of the generalization and robustness ability with \change{unseen external domains (ED-1, ED-2)}. Data leakage is carefully considered to be avoided during training and testing. All the results reported for the \change{testing} domains are derived from testing on unseen data. Specifically, for the Retina segmentation experiments, the model is saved by considering its performance on the test-split of SD. Besides, only the train-split of TD is used for amplitude fusion during the curriculum-style training. The test split of TD and the \change{external domains (ED-1, ED-2)} are used for model evaluation. For the Nuclei segmentation task, to avoid the training and evaluation bias due to the typically small dataset size, we perform the 5-fold (folds are split based on human organs) cross-validation for training and report the average result on the left-out fold of the source domain. Similarly, the test-split of TD and the entire \change{ED} are used in performance assessment.

Regarding robustness evaluation, we test the best model derived from each method with different corruption types and levels of synthetic Retina datasets. The performance is compared in two aspects, i.e., robustness under 15 corruption types on average of 5 corruption levels and robustness under 5 corruption levels on average of 15 corruption types. 

\subsection{Results Analysis}

\begin{figure*}[!ht]
\includegraphics[width=0.72\textwidth]{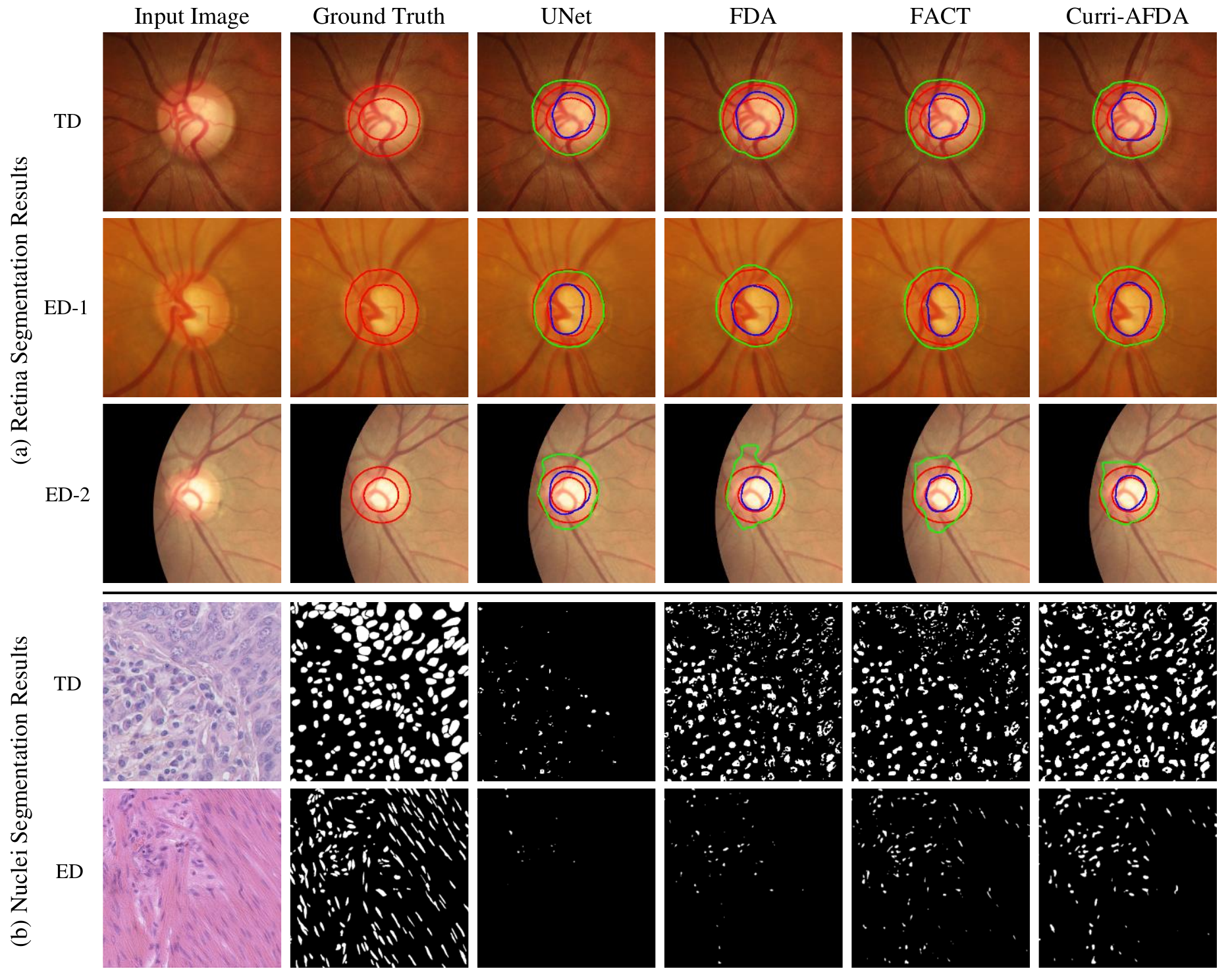}
\centering
\caption{\textbf{Qualitative comparison on the results of different methods with UNet~\cite{ronneberger2015unet} backbone for (a) Retina segmentation and (b) Nuclei segmentation.} Each row demonstrates the segmentation results of different methods compared with the ground truth for the \change{testing images}. In (a), the blue and green contours indicate the boundaries of the optic cups and optic discs, respectively, while the red contours are the ground truths. The boundaries of both classes obtained by our Curri-AFDA are closer to the ground truths. In (b), more nuclei can be segmented out by our method for both \change{testing images}.}
\label{fig:maskoverlay}
\end{figure*}

The overall quantitative and qualitative results are shown in Table.~\ref{tab:overall_result} and Fig.~\ref{fig:maskoverlay}. The results suggest the superior performance of our method Curri-AFDA in both domain-adaptive segmentation tasks compared with other methods, i.e., vanilla UNet~\cite{ronneberger2015unet}, vanilla Swin-UNet~\cite{cao2023swin}, ADVENT~\cite{vu2019advent}, FDA~\cite{yang2020fda} and FACT~\cite{xu2021fourier}. 

\subsubsection{Retina Segmentation}
For the Retina Segmentation task, \change{on the target domain (TD)}, our Curri-AFDA achieves the DSC improvement of $0.96\%$ and $5.30\%$ against the best result from the other methods with UNet~\cite{ronneberger2015unet} and Swin-UNet~\cite{cao2023swin} as backbones, respectively. This shows the outstanding adaptation performance of our approach. When comparing the generalization and robustness performance on the unseen \change{external domains, i.e., ED-1 and ED-2}, our method also achieves the best result in most cases. Note that with the Swin-UNet~\cite{cao2023swin} backbone, our method yields a bit lower result than the best one from other methods. We attribute this to the relatively small dataset size. However, on average of all three testing datasets, our Curri-AFDA can improve the DSC performance by $1.78\%$ and $2.68\%$ for the two backbones, showing the superior generalization and robustness ability. As demonstrated in Fig.~\ref{fig:maskoverlay}(a), more accurate segmentation masks and boundaries can be obtained with our method. 

\begin{table*}[!htbp]
\centering
\caption{\textbf{Robustness performance of our Curri-AFDA and other methods on corrupted retina data under various types of corruption.} Results are obtained by averaging the performance under five severity levels for each corruption type. Our Curri-AFDA outperforms other methods by a large margin for most of the corruption types. The best DSC (\%) results are highlighted in bold.}
\label{tab:robustness}
\resizebox{\textwidth}{!}{
\begin{tabular}{c|ccc|cccc|cccc|cccc|c} 
\toprule
\multirow{2}{*}{\textbf{Methods}} & \multicolumn{3}{c|}{\textbf{Noise}}              & \multicolumn{4}{c|}{\textbf{Blur}}                                & \multicolumn{4}{c|}{\textbf{Weather}}                             & \multicolumn{4}{c|}{\textbf{Digital}}                            & \multirow{2}{*}{\textbf{Mean}}  \\ 
\cline{2-16}
                                  & Gauss.         & Shot           & Impulse        & Defocus        & Glass          & Motion         & Zoom           & Snow           & Frost          & Fog            & Bright         & Contrast       & Elastic        & Pixel          & JPEG           &                                 \\ 
\hline
UNet~\cite{ronneberger2015unet}                              & 70.77          & 70.24          & 70.49          & 70.05          & 70.09          & 69.19          & 68.39          & 68.43          & 65.45          & 54.78          & 68.78          & 40.19          & 64.49          & 70.28          & 69.74          & 66.09                           \\ 
+ FDA~\cite{yang2020fda}                              & 76.30          & 76.44          & 71.07          & 78.34          & 78.98          & 77.06          & 76.20          & 60.20          & 68.11          & \textbf{64.74} & 77.55          & 35.45          & 72.37          & 79.33          & 76.95          & 71.27                           \\ 
+ FACT~\cite{xu2021fourier}                             & 74.55          & 75.08          & 73.30          & 78.03          & 78.44          & 76.17          & 76.21          & 60.20          & 65.27          & 59.95          & 73.53          & \textbf{56.56} & 71.28          & 78.14          & 75.61          & 71.49                           \\ 
\hline
+ Curri-AFDA (Ours)                         & \textbf{78.87} & \textbf{79.45} & \textbf{78.24} & \textbf{80.43} & \textbf{80.65} & \textbf{78.33} & \textbf{77.81} & \textbf{71.04} & \textbf{75.41} & 59.89          & \textbf{77.79} & 52.59          & \textbf{73.03} & \textbf{80.40} & \textbf{79.33} & \textbf{74.88}                  \\
\bottomrule
\end{tabular}}
\label{table:robusness}
\end{table*}

\subsubsection{Nuclei Segmentation}
For the Nuclei Segmentation task which is much harder due to multiple tiny instances with uncertain positions, similar conclusions can also be drawn that our curriculum-based approach outperforms other methods regarding adaptation, generalization, and robustness, with much more significant gains. Although the performance on ED with Swin-UNet~\cite{cao2023swin} is a bit lower than some other methods due to the smaller dataset size, the overall DSC gains are $8.50\%$ and $4.29\%$ on the testing data with UNet~\cite{ronneberger2015unet} and Swin-UNet~\cite{cao2023swin} backbones. Fig.~\ref{fig:maskoverlay}(b) shows the qualitative comparison of different methods for Nuclei segmentation. Our method performs better in such a segmentation task to recognize more nuclei.

\begin{figure}[!t]
\includegraphics[width=0.92\columnwidth]{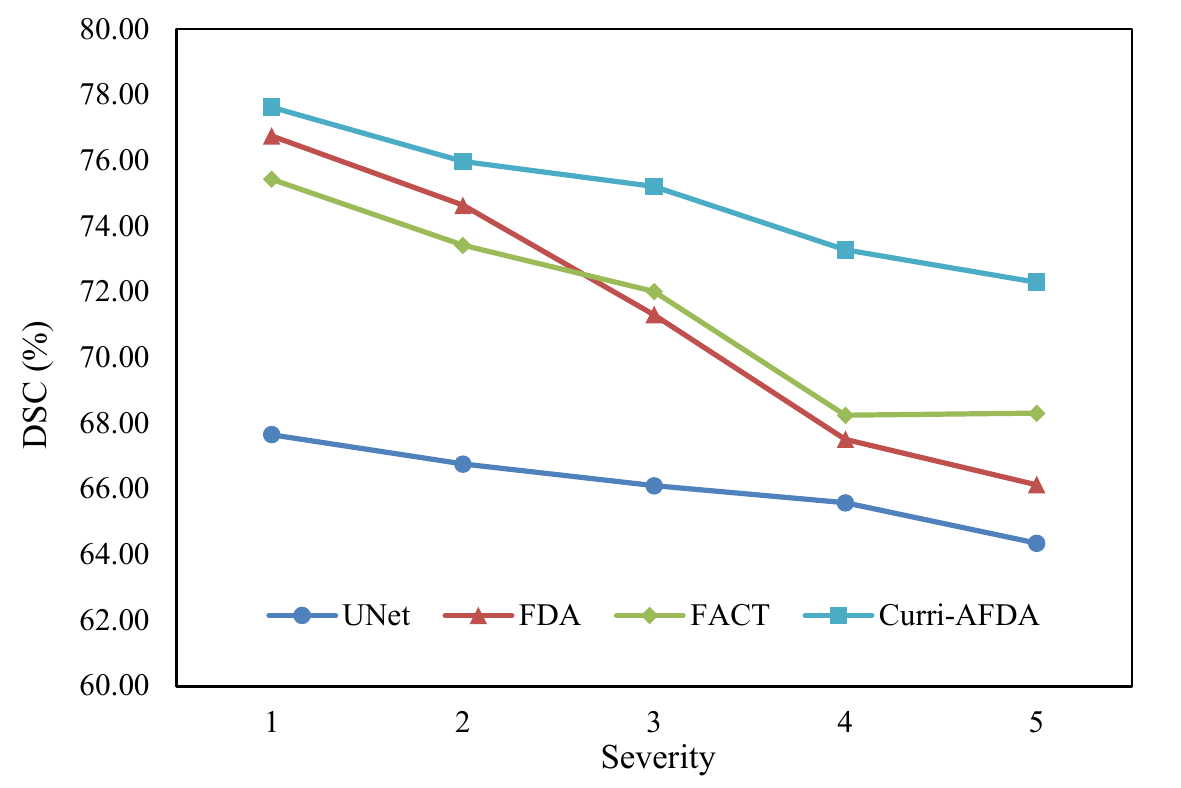}
\centering
\caption{\textbf{Robustness comparison of different methods on the synthetic retina data under growing severity levels of corruption on average of different corruption types.} Our Curri-AFDA is more robust to preserve higher and stabler performance.}
\label{fig:robustness_severity}
\end{figure}
\subsubsection{Robustness Analysis} 
A more robust model is reflected in the fact that it can still maintain higher performance when exposed to corrupted images under different corruption types and increased corruption severity levels~\cite{hendrycks2019benchmarking}. On the one hand, as shown in Table.~\ref{table:robusness}, our Curri-AFDA yields higher performance against most of the corruption types compared with other methods. The overall average DSC of our Curri-AFDA surpasses the best result of other methods by $3.39\%$. On the other hand, Fig.~\ref{fig:robustness_severity} illustrates that our Curri-AFDA maintains superior performance under increasing severity levels while the performance of other approaches degrades dramatically, especially in comparison to the other two Fourier-based approaches.

In summary, the proposed framework of curriculum-based amplitude fusion and chained augmentation mixing allows the model to explore and learn a broader feature representation space. The results of extensive experiments and evaluation on multiple domains indicate that our Curri-AFDA is generic and capable of achieving superior adaptation, generalization, and robustness performance compared with other methods.

\section{Ablation Study}
\label{sec:ablation}

\subsection{Decomposition Analysis of Each Module}
\begin{figure}[!t]
\includegraphics[width=0.92\columnwidth]{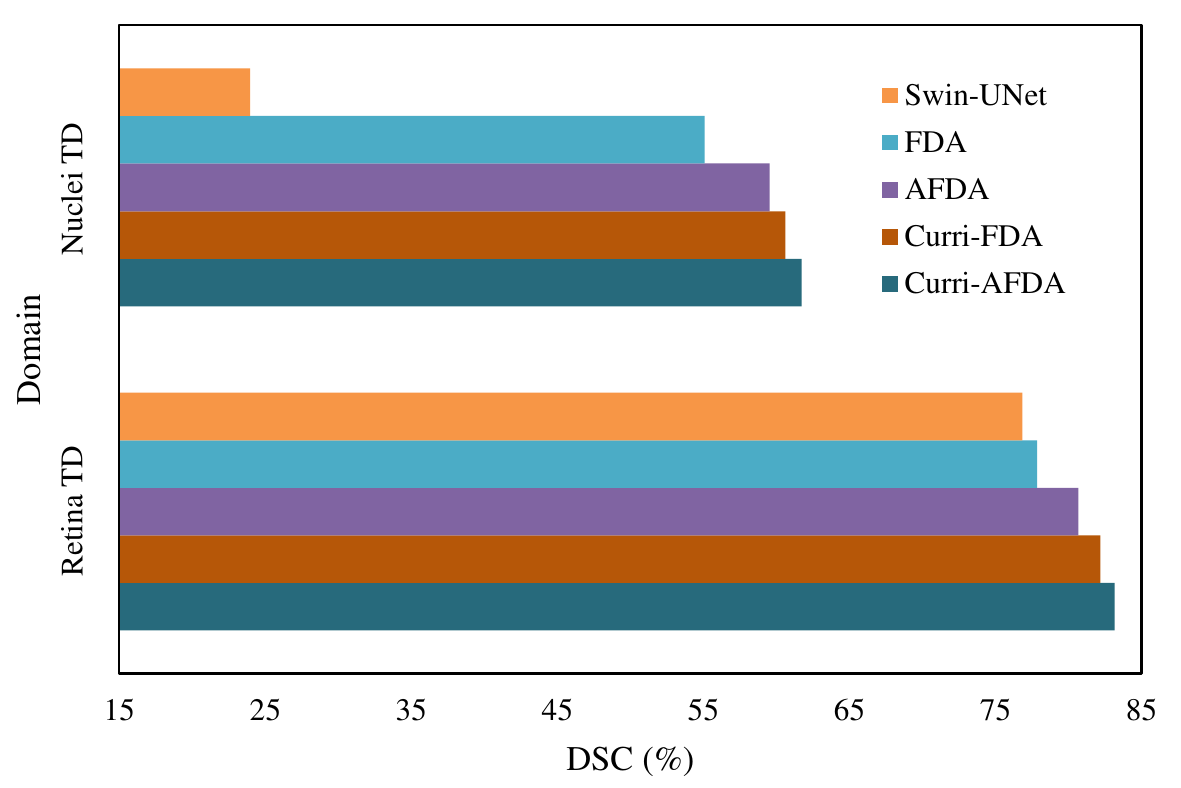}
\centering
\caption{\textbf{Ablation comparison of our method Curri-AFDA with the Swin-UNet~\cite{cao2023swin} backbone on two target domains of Retina and Nuclei segmentation.} Consistently improved results demonstrate the efficacy of each module in our proposal.}
\label{fig:aug_curri}
\end{figure}

As shown in Fig.~\ref{fig:proposal}, our proposal mainly consists of three modules, i.e., Amplitude Fusion, Curriculum Learning and Augmentation Mixing. Here we decouple these modules and compare the performance with Vanilla Swin-UNet~\cite{cao2023swin} (without adaptation method), FDA~\cite{yang2020fda}, AFDA (FDA~\cite{yang2020fda} with Augmentation Mixing) and Curri-FDA (FDA~\cite{yang2020fda} with our curriculum strategy). As shown in Fig.~\ref{fig:aug_curri}, the three modules can consistently improve the performance and the integration of them, i.e., our Curri-AFDA, yields the best results on both target domains of Retina and Nuclei segmentation.

\subsection{Curriculum Vs. Anti-Curriculum Vs. Random}

Depending on a comprehensive understanding of the training data and task, the design of the curriculum is of vital significance in the utilization of Curriculum Learning. For the domain adaption and generalization task, amplitude fusion of images in the frequency space can help mitigate the variance between different domains. We take advantage of this property and establish an effective curriculum-based training framework. Specifically, in our hypothesis, the amplitude scaling coefficient $\beta$ controls the amount of target domain information to be transferred to the source domain. The scheduled increment of $\beta$, i.e., the $\beta_c$ in~\eqref{eq:lin_sche}, is the core idea in our proposed curriculum for the domain-adaptive segmentation task. By gradually increasing the amount of the mutated low-frequency amplitude components in the source and target domain data, the generated training samples carry more domain-invariant information and thus enable the model to be more generalizable.

\begin{figure}[!t]
\includegraphics[width=0.92\columnwidth]{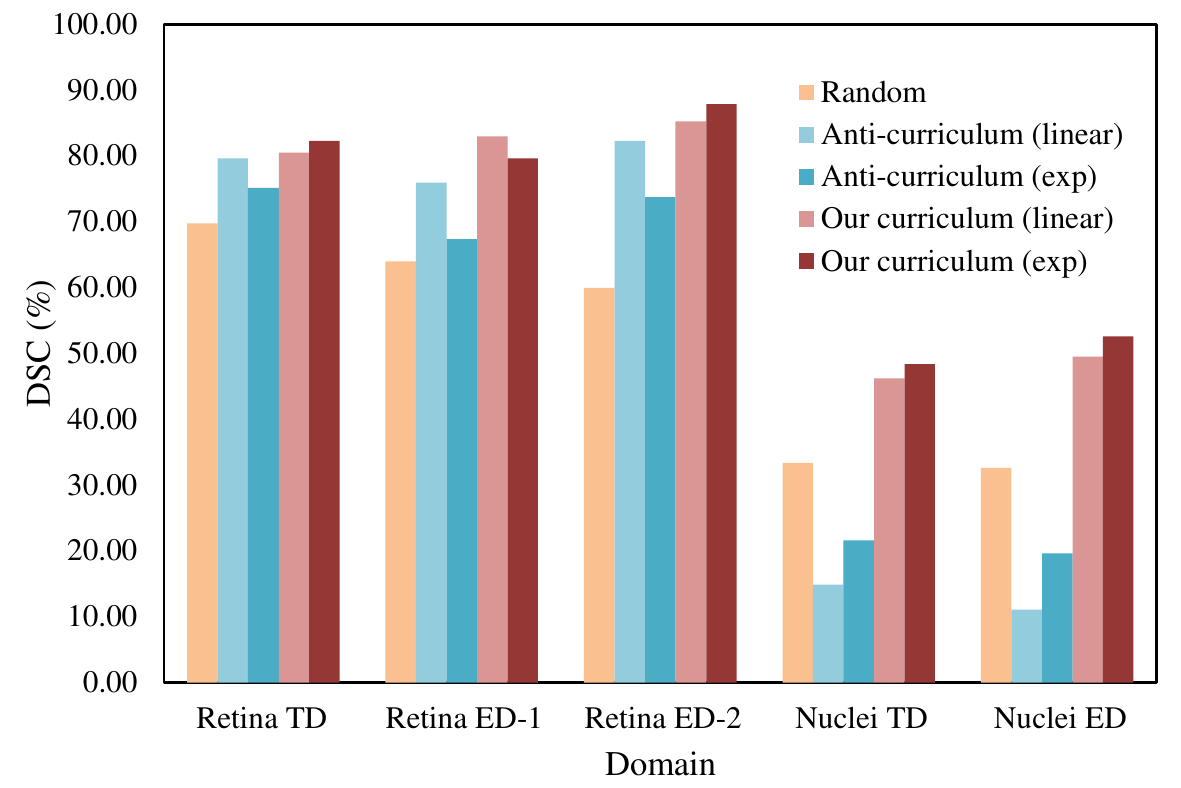}
\centering
\caption{\textbf{Results of different curriculum designs on \change{testing} domains of Retina segmentation and Nuclei segmentation with UNet~\cite{ronneberger2015unet} backbone.} Our curriculum can always yield better performance, especially on the nuclei datasets.}
\label{fig:curriculum_strategies}
\end{figure}

Apart from our curriculum, we notice that in some works~\cite{fan2018learning, wang2020learning}, the best curriculum is reported as the opposite of conventional curriculum learning, i.e., ``harder to easier''. Therefore, we also conduct experiments with another two curriculum designs, i.e., the anti-curriculum in which $\beta_c$ gradually decreases and the random-curriculum in which $\beta_c$ is randomly sampled in the range $[0, \beta_{opt}]$. For a comprehensive comparison, both the linear and exponential scheduling functions of $\beta_c$ are evaluated and reported. As illustrated in Fig.~\ref{fig:curriculum_strategies}, our curriculum design yields better results on all \change{testing} domains for both tasks, especially for the Nuclei segmentation task.

\subsection{Sensitivity to Epoch Ratio} \label{sec:abl_er}

Epoch ratio ($r_e$ in~\eqref{eq:lin_sche}) controls the duration of applying our curriculum strategy in the whole training process and \change{affects the changing rate of the amplitude scaling coefficient $\beta_c$. This further characterizes different learning speeds of cross-domain information.}
We present the ablation study on the Retina segmentation task to compare the performance of our method Curri-AFDA with FDA~\cite{yang2020fda} under different epoch ratios. The exponential function is adopted to update the amplitude scaling coefficient. We take the constant FDA~\cite{yang2020fda} results as a reference for its irrelevance to the epoch ratio.

\begin{figure}[!t]
\includegraphics[width=0.92\columnwidth]{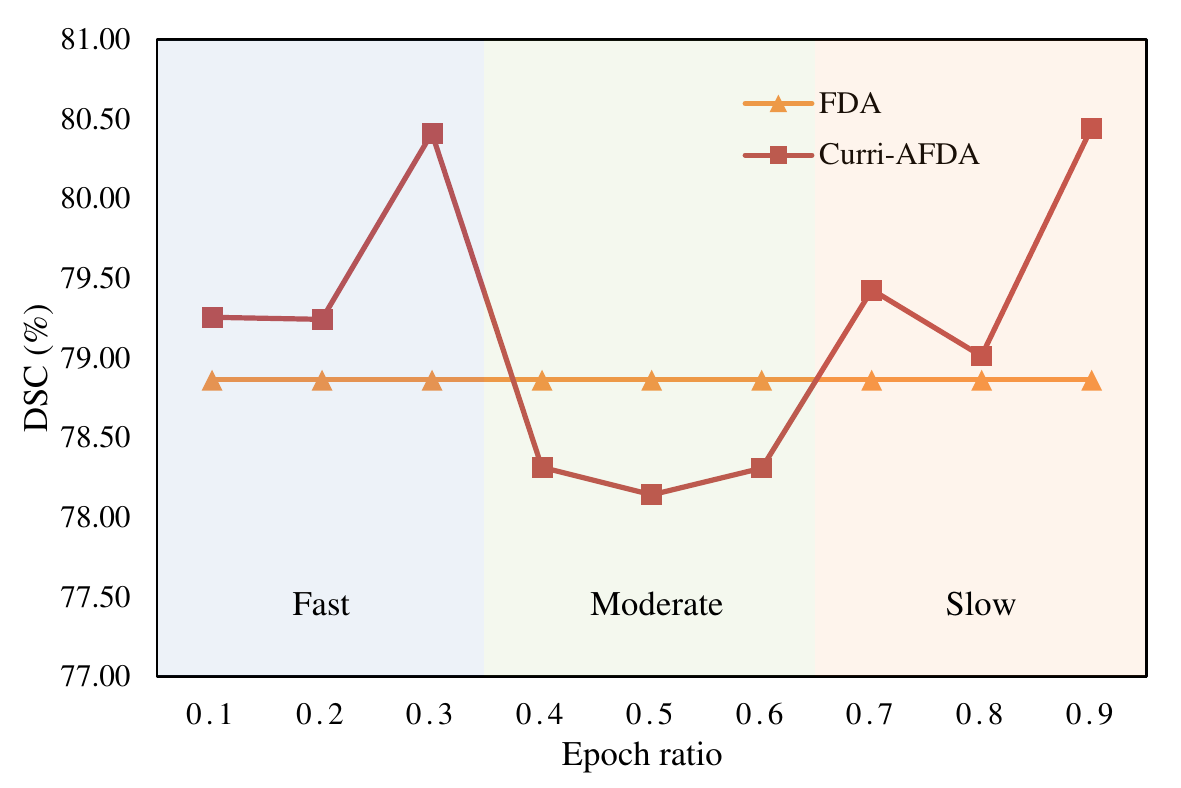}
\centering
\caption{\change{\textbf{Results of tuning epoch ratios.} 
Faster and slower learning speeds characterized by smaller and larger epoch ratios are more likely to enhance performance.}}
\label{fig:analysis_epoch_ratio}
\end{figure}

\change{In our curriculum-based domain-adaptive segmentation task, smaller or larger epoch ratios result in quicker or more gradual exposure of TD information. This can let the model learn cross-domain information \textbf{faster} within earlier epochs or \textbf{slower} until later epochs. As shown in Fig.~\ref{fig:analysis_epoch_ratio}, these two learning speeds are more likely to enhance the model performance than a moderate one. Such behavior aligns with the generic Curriculum Leaning theory~\cite{wu2021when}, which suggests that models exhibit improved performance by either learning the challenging task \textbf{faster} within earlier epochs or \textbf{slower} until later epochs, rather than adopting a moderate learning pace. The findings in this ablation study prioritize initializing the epoch ratio with a smaller or larger value to achieve optimal results.}

\subsection{Linear and Exponential Scheduling Functions}

\begin{table}[!t]
  \centering
  \caption{\textbf{Comparison of linear and exponential scheduling functions under various epoch ratios.} The exponential function provides a higher average DSC while the linear function gives stabler performance under different epoch ratios with a lower standard deviation.}
  \resizebox{\columnwidth}{!}{
    \begin{tabular}{ccccccccccc}
    \toprule
    \multirow{1}[4]{*}{\textbf{Scheduler}} & \multicolumn{8}{c}{\textbf{Epoch ratio} \boldmath $r_e$}                              & \multirow{1}[4]{*}{\textbf{Mean}} & \multirow{1}[4]{*}{\textbf{STD}} \\
\cline{2-9}          & 0.2   & 0.3   & 0.4   & 0.5   & 0.6   & 0.7   & 0.8   & 0.9   &       &  \\
    \hline
    Linear & \textbf{46.02} & 42.81 & \textbf{45.14} & 41.86 & \textbf{43.79} & 39.77 & 41.05 & \textbf{42.27} & 42.84 & \textbf{2.08} \\
    Exponential   & 38.87 & \textbf{45.79} & 44.40 & \textbf{43.64} & 41.75 & \textbf{47.60} & \textbf{50.61} & 40.05 & \textbf{44.09} & 3.91 \\
    \bottomrule
    \end{tabular}%
    }
  \label{tab:lin_exp_comp}%
\end{table}%

We further explore the scheduling functions of the amplitude scaling coefficient $\beta$ in our curriculum. The way to update $\beta$ is one of the major considerations in designing our approaches. Specifically, we implement the predefined linear and exponential functions to update $\beta$ in each training epoch. As shown in Fig.~\ref{fig:lin_exp}, the key difference between them is that for a fixed epoch ratio, the exponential function yields a variable changing rate of $\beta$. In contrast, the linear function provides a constant one throughout the curriculum. Here we investigate the effect of these two scheduling functions with experiments on the Nuclei segmentation task with UNet~\cite{ronneberger2015unet} backbone. 

\change{As outlined in Section.~\ref{sec:abl_er}, smaller or larger epoch ratios are more likely to yield improved results than intermediate ones. The results in Table.~\ref{tab:lin_exp_comp} further substantiate this conclusion by showing that the optimal performance of the linear and exponential schedulers are achieved with epoch ratios of 0.2 and 0.8, respectively. These values fall within the suggested feasible range of epoch ratios.
Upon examining the results presented in Table.~\ref{tab:lin_exp_comp} within the suggested ranges of epoch ratios, we can observe that the exponential scheduler outperforms the linear scheduler in more cases. Furthermore, we note that the exponential scheduler achieves a higher average performance across all epoch ratios. These findings indicate that the exponential scheduler is more reliable in providing favorable outcomes than the linear scheduler with our method.}

\subsection{Efficacy for Medical Image Classification}
\begin{table}[!t]
\centering
\caption{\textbf{Statistics of skin lesion classification datasets.} Over 10 thousand samples are included in four medical domains.}
\label{tab:skin_data}
\resizebox{0.82\linewidth}{!}{%
\begin{tabular}{ccccc} 
\toprule
\multicolumn{1}{c}{\multirow{2}{*}{\textbf{Domains}}} & \multicolumn{3}{c}{\textbf{Lesion Types}} & \multicolumn{1}{c}{\multirow{2}{*}{\textbf{Total}}}  \\ 
\cline{2-4}
\multicolumn{1}{c}{}                                 & Nevus & Benign Keratosis & Melanoma       & \multicolumn{1}{c}{}                                 \\ 
\hline
SD                                                      & 1372  & 254              & 374            & 2000                                                  \\
\change{TD}                                                      & 803   & 490              & 342            & 1635                                                  \\ 
\change{ED-1}                                                      & 1832  & 475              & 680            & 2987                                                  \\ 
\change{ED-2}                                                      & 3720  & 124              & 24             & 3868                                                  \\ 
\bottomrule
\end{tabular}
}
\end{table}

\begin{table}[!t]
  \centering
  \caption{\textbf{Quantitative results of skin lesion classification with Swin-Transformer~\cite{liu2021swin}}. The f1 score (\%) results and the overall performance on the \change{testing} domains are reported. The best results are in bold and the runner-up results are underlined.}
  \resizebox{\linewidth}{!}{%
    \begin{tabular}{c|ccccc}
    \toprule
    \multicolumn{1}{c|}{\textbf{Methods}} & \boldmath $\beta$ & \multicolumn{1}{c}{\textbf{\change{TD}}} & \multicolumn{1}{c}{\textbf{\change{ED-1}}} & \multicolumn{1}{c}{\textbf{\change{ED-2}}} & \textbf{Mean$\pm$STD} \\
    \hline
    \multicolumn{1}{c|}{Swin-Transformer~\cite{liu2021swin}} & N.A.  & 36.27 & 49.89 & 29.07 & 38.41$\pm$8.64 \\
    \multicolumn{1}{c|}{+ FDA~\cite{yang2020fda}} & \multicolumn{1}{c}{0.06} & 40.33 & 49.07 & 39.19 & 42.86$\pm$4.41 \\
    \multicolumn{1}{c|}{+ FACT~\cite{xu2021fourier}} & \multicolumn{1}{c}{1} & 39.73 & \underline{54.01} & \textbf{47.64} & \underline{47.13}$\pm$5.84 \\
    \hline
    \multicolumn{1}{c|}{\multirow{2}[1]{*}{+ Curri-AFDA (Ours)}} & linear (0 to 0.06) & \underline{41.46} & \textbf{56.58} & 44.70 & \textbf{47.58}$\pm$6.50 \\
          & exp (0 to 0.06) & \textbf{41.49} & 47.69 & \underline{47.05} & 45.41$\pm$2.78 \\
    \bottomrule
    \end{tabular}%
    }
  \label{tab:skin_res}%
\end{table}%

Besides segmentation, image classification is also fundamentally demanding in medical applications. To evaluate the efficacy of our Curri-AFDA for medical image classification, we utilize a collection of skin lesion datasets with thousands of samples released by PRR-FL~\cite{chen2021personalized}. The datasets have four medical domains and are annotated with three skin lesion types, i.e., Nevus, Benign Keratosis, and Melanoma. Following their dataset splits as shown in Table.~\ref{tab:skin_data}, we conduct the classification experiments with the Swin-Transformer~\cite{liu2021swin} as the backbone. The f1 score is adopted as the evaluation metric to reveal a more comprehensive comparison of the unbalanced datasets. The optimal amplitude scaling coefficient $\beta_{opt}$ and the weighting coefficient $\alpha$ are 0.06 and 0.7, respectively.

As shown in Table.~\ref{tab:skin_res}, our method yields the best overall result of f1 score on the skin lesion datasets. This demonstrates that our methods are also supportive of the domain-adaptive medical image classification task in addition to segmentation.

\section{Conclusion and Discussion}
\label{sec:conclusion}

This work proposes the Curriculum-based Augmented Fourier Domain Adaptation (Curri-AFDA) and proves to achieve superior adaptation, generalization, and robustness performance for medical image segmentation. Specifically, we design a novel curriculum strategy to progressively transfer amplitude information in the Fourier space from the target domain to the source domain to mitigate domain gaps and incorporate the chained augmentation mixing to further improve the generalization and robustness ability. Our method is naturally modality-independent due to its independence on any particular properties of the imaging modality. Without additional trainable parameters, extensive experiments on two segmentation tasks with multiple-domain datasets of two image modalities demonstrate the efficacy of our method on top of both the classical CNN (UNet~\cite{ronneberger2015unet}) and recent transformer (Swin-UNet~\cite{cao2023swin}) architectures. Specially, we consider the crucial yet rarely explored topic in medical image analysis, i.e., the robustness performance with the synthetic dataset generated by different types and levels of corruptions, and also observe the superior results of our method. Additionally, our method can also contribute to medical image classification besides segmentation, indicating its potential for broader medical applications.

Future research may focus on designing more flexible and automatic scheduling functions to update the amplitude scaling coefficient which adjusts the amplitude fusion area. In addition, the weighting coefficient which controls the merging ratio between images can also be involved when designing the curriculum strategy. Besides the training-time Domain Adaptation, test-time Domain Adaptation~\cite{karani2021test, sun2020test} is also worth to be explored by integrating the Fourier-based cross-domain information fusion and the chained augmentation mixing.

\bibliographystyle{IEEEtran}
\bibliography{mybib}
 

\begin{IEEEbiography}[{\includegraphics[width=1in,height=1.25in,clip,keepaspectratio]{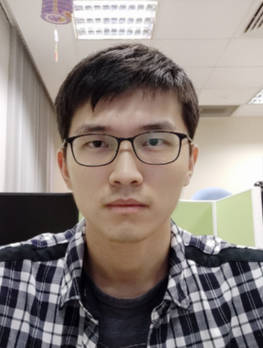}}]{An Wang} 
received his B.Eng. degree in Information Engineering from Soochow University, SuZhou, China, in 2018, and his M.Sc. degree in Electrical Engineering from National University of Singapore, Singapore, in 2019. He is currently pursuing the Ph.D. degree at Medical Mechatronics Lab of the Department of Electronic Engineering, The Chinese University of Hong Kong, supervised
by Prof. Hongliang Ren. His research focuses on efficient medical image analysis and computer-assisted intervention.
\end{IEEEbiography}

\vspace{-50 pt} 
\begin{IEEEbiography}[{\includegraphics[width=1in,height=1.25in,clip,keepaspectratio]{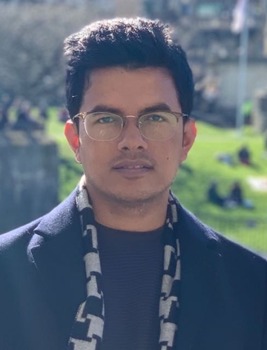}}]{Mobarakol Islam} 
received his Ph.D. degree from NUS Graduate School for Integrative Sciences and Engineering (NGS), National University of Singapore in Dec 2019. He is now a senior research fellow at the Department of Medical Physics and Biomedical Engineering, University College London, working with Dr. Matt Clarkson in WEISS. Before that, he was a postdoctoral research associate at the Department of Computing, Imperial College London, under the supervision of Dr. Ben Glocker in BioMedIA Lab. His research focuses on enhancing deep neural network robustness, fairness, and reliability using calibration, uncertainty, and causality to improve image-guided disease diagnosis and intervention.
\end{IEEEbiography}
\vspace{-50 pt}

\begin{IEEEbiography}[{\includegraphics[width=1in,height=1.25in,clip,keepaspectratio]{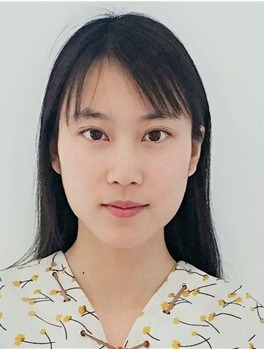}}]{Mengya Xu}
received her B.Eng. degree in information engineering from the Soochow University, SuZhou, China, in 2018, and her M.Sc. degree in Electrical Engineering from  National University of Singapore, Singapore, in 2019.
She is currently pursuing the Ph.D. degree at the Department of Biomedical Engineering, National University of Singapore, Singapore, supervised
by Prof. Hongliang Ren.
Her research focuses on vision-language multimodality-based surgical scene understanding. 
\end{IEEEbiography}
\vspace{-50 pt}

\begin{IEEEbiography}[{\includegraphics[width=1in,height=1.25in,clip,keepaspectratio]{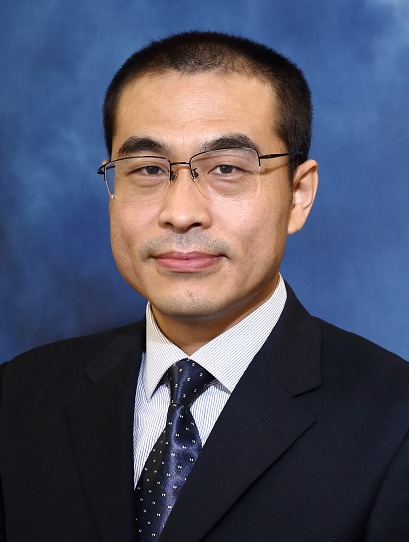}}]{Hongliang Ren}
received his Ph.D. in Electronic Engineering (Specialized in Biomedical Engineering) from The Chinese University of Hong Kong (CUHK) in 2008. He has been navigating his academic journey through Chinese University of Hong Kong, UC Berkeley, Johns Hopkins University, Children’s Hospital Boston, Harvard Medical School, Children’s National Medical Center, United States, and National University of Singapore. He serves as an Associate Editor for IEEE Transactions on Automation Science \& Engineering (T-ASE) and Medical \& Biological Engineering \& Computing (MBEC). He has served as an active organizer and contributor on the committees of numerous robotics conferences, including a variety of roles in the flagship IEEE Conf. on Robotics and Automation (ICRA), IEEE Conf. on Intelligent Robots and Systems (IROS), as well as other domain conferences such as ROBIO/BIOROB/ICIA. He served as publicity chair for ICRA 2017, concurrently as Organizing Chair for ICRA 2017 workshop on Surgical Robots, and video chair for ICRA 2021. He has delivered numerous invited keynotes/talks at flagship conferences/workshops at ICRA/IROS/ROBIO/ICIA. He is the recipient of IFMBE/IAMBE Early Career Award 2018, Interstellar Early Career Investigator Award 2018, and ICBHI Young Investigator Award 2019. He is also the recipient of numerous international conference awards, including Best Conference Paper Awards at IEEE ROBIO 2019, IEEE RCAR 2016, IEEE CCECE 2015, IEEE Cyber 2014, and IEEE ROBIO 2013. His research is mainly at Biorobotics \& intelligent systems, medical mechatronics, continuum, and soft flexible robots and sensors, multisensory perception, learning and control in image-guided procedures, deployable motion generation, compliance modulation/sensing, cooperative and context-aware sensors/actuators in human environments, robotic surgery, flexible robotics and machine artificial intelligence.
\end{IEEEbiography}

\end{document}